\documentclass[prl,letterpaper,twocolumn,showpacs,superscriptaddress,amsmath,amsfonts,amssymb,floatfix,preprintnumbers]{revtex4}
\pdfoutput=1
\usepackage[T1]{fontenc}\usepackage[latin1]{inputenc}
\usepackage{dcolumn,graphicx,color,hvfloat,booktabs,microtype,afterpage} 
\usepackage[charter,greekuppercase=italicized]{mathdesign}
\renewcommand{\figurename}{Fig.}
\renewcommand{\tablename}{Table}
\makeatletter\renewcommand{\fnum@figure}[1]{\figurename~\thefigure~(color online).}\makeatother
\makeatletter\renewcommand{\fnum@table}[1]{\tablename~\thetable.}\makeatother

\usepackage[colorlinks,plainpages=false,linkcolor=blue,urlcolor=blue,citecolor=blue,pdfpagemode=UseNone,pdfstartview=FitBH]{hyperref}
\newcount\hh \newcount\mm
\hh=\time \divide\hh by 60
\mm=\hh \multiply\mm by 60 \mm=-\mm
\advance\mm by \time
\def\now{\number\hh:\ifnum\mm<10{}0\fi\number\mm}

\newcommand{\vect}[1]{\mathbf{#1}}
\newcommand{\half}{\frac{1}{\protect\raisebox{0.8pt}{\scriptsize 2}}}
\newcommand{\quarter}{\frac{1}{\protect\raisebox{0.8pt}{\scriptsize 4}}}
\newcommand{\quarterIII}{\frac{3}{\protect\raisebox{0.8pt}{\scriptsize 4}}}

\begin{document}

\makeatletter\renewcommand{\ps@plain}{%
\def\@evenhead{\hfill\itshape\rightmark}%
\def\@oddhead{\itshape\leftmark\hfill}%
\renewcommand{\@evenfoot}{\hfill\small{--~\thepage~--}\hfill}%
\renewcommand{\@oddfoot}{\hfill\small{--~\thepage~--}\hfill}%
}\makeatother\pagestyle{plain}

\title{Reciprocal-space structure and dispersion of the magnetic resonant mode\\in the superconducting phase of Rb$_x$Fe$_{2-\!y}$Se$_2$ single crystals}

\author{G.~Friemel}
\affiliation{Max-Planck-Institut für Festkörperforschung, Heisenbergstraße 1, 70569 Stuttgart, Germany}
\author{J.\,T.~Park}
\affiliation{Max-Planck-Institut für Festkörperforschung, Heisenbergstraße 1, 70569 Stuttgart, Germany}
\author{T.~A.~Maier}
\affiliation{Computer Science and Mathematics Division and Center for Nanophase Materials Sciences, Oak Ridge National Lab, Oak Ridge, TN~37831, USA}
\author{V.~Tsurkan}
\affiliation{Center for Electronic Correlations and Magnetism, Institute of Physics, Augsburg University, 86135 Augsburg, Germany}
\affiliation{Institute of Applied Physics, Academy of Sciences of Moldova, MD 2028, Chisinau, Republic of Moldova}
\author{Yuan~Li}
\affiliation{Max-Planck-Institut für Festkörperforschung, Heisenbergstraße 1, 70569 Stuttgart, Germany}
\author{J.~Deisenhofer}
\affiliation{Center for Electronic Correlations and Magnetism, Institute of Physics, Augsburg University, 86135 Augsburg, Germany}
\author{H.-A.~Krug~von~Nidda}
\affiliation{Center for Electronic Correlations and Magnetism, Institute of Physics, Augsburg University, 86135 Augsburg, Germany}
\author{A.~Loidl}
\affiliation{Center for Electronic Correlations and Magnetism, Institute of Physics, Augsburg University, 86135 Augsburg, Germany}
\author{A.~Ivanov}
\affiliation{Institut Laue-Langevin, 6 rue Jules Horowitz, 38042 Grenoble Cedex 9, France}
\author{B.~Keimer}
\affiliation{Max-Planck-Institut für Festkörperforschung, Heisenbergstraße 1, 70569 Stuttgart, Germany}
\author{D.\,S.\,Inosov{\hspace{6pt}$^,$\large\hyperref[CorrAuthor]{*}\hspace*{-13pt}}}
\affiliation{Max-Planck-Institut für Festkörperforschung, Heisenbergstraße 1, 70569 Stuttgart, Germany}

\begin{abstract}
Inelastic neutron scattering is employed to study the reciprocal-space structure and dispersion of magnetic excitations in the normal and superconducting states of single-crystalline Rb$_{0.8}$Fe$_{1.6}$Se$_2$. We show that the recently discovered magnetic resonant mode in this compound has a quasi-two-dimensional character, similar to overdoped iron-pnictide superconductors. Moreover, it has a rich in-plane structure that is dominated by four elliptical peaks, symmetrically surrounding the Brillouin zone corner, without $\sqrt{5}\times\!\sqrt{5}$ reconstruction. We also present evidence for the dispersion of the resonance peak, as its position in momentum space depends on energy. Comparison of our findings with the results of band structure calculations provides strong support for the itinerant origin of the observed signal. It can be traced back to the nesting of electronlike Fermi pockets in the doped metallic phase of the sample in the absence of iron-vacancy ordering.
\end{abstract}


\maketitle\thispagestyle{plain}\enlargethispage{3pt}

The newly discovered iron selenide superconductors \textit{A}$_x$Fe$_{2-y}$Se$_2$ ($A$\,=\,K,\,Rb,\,Cs) \cite{Discovery} became famous for their relatively high critical temperature, $T_\text{c}=32$\,K, observed concurrently with a strong antiferromagnetic (AFM) order that persists far above room temperature \cite{Shermadini11, Yan11}. However, an ordered moment as large as $3.3\,\mu_\text{B}/\text{Fe}$ \cite{order} renders microscopic coexistence \cite{ZhangXiao11} doubtful. The superconducting (SC) phase usually appears in samples close to the 2:4:5 stoichiometry \cite{Yan11, Tsurkan11}, which is at the same time the optimal composition for the ordering of Fe vacancies into a $\sqrt{5}\times\!\sqrt{5}$ superstructure, grouping the occupied iron sites in plaquettes of four ferromagnetically aligned moments. On the one hand, experiments \cite{Yan11, stm, Charnukha11} and band structure calculations \cite{YanGao11, CaoDai11} suggest this superstructure to be insulating. On the other hand, angle-resolved photoelectron spectroscopy (ARPES) revealed a Fermi surface (FS) dominated by a large electron pocket at the $M$ point \cite{Gap1,Gap2}. Recent reports reconcile these seemingly contradictory findings by the observation of several coexisting phases, seen in transmission electron microscopy (TEM) \cite{WangSong11, SongWang11, YuanDong11, KazakovAbakumov11}, scanning tunneling microscopy (STM) \cite{stm}, x-ray diffraction \cite{xray}, ARPES \cite{ChenXu11}, magnetization measurements \cite{ShenZeng11}, muon-spin rotation ($\mu$SR) \cite{ShermadiniLuetkens11}, M\"{o}ssbauer \cite{KsenofontovWortmann11} and optical \cite{Charnukha11} spectroscopies. While STM studies observed the SC gap on a vacancy-free surface \cite{stm}, TEM measurements suggested that the second phase is an iron-vacancy disordered state \cite{WangSong11, SongWang11}. The phase separation scenario clearly needs more clarification in terms of the structure and stoichiometry of the SC phase for a consistent understanding of these observations.

In iron pnictides, it is established that the SC order parameter changes its sign between the hole- and electronlike sheets of the FS \cite{spm, LumsdenChristianson09, InosovPark10}. Despite the absence or strong reduction of the hole Fermi pocket in iron selenides \cite{LDA}, different kinds of a sign-changing gap have also been suggested \cite{Mazin11, Maier11, DasBalatsky11}. The recent finding of a magnetic resonant mode in the low-energy spin-excitation spectrum of Rb$_{0.8}$Fe$_{1.6}$Se$_2$ below $T_\text{c}$ \cite{ParkFriemel11} supports these unconventional pairing scenarios. Its wave vector, $\vect{Q}=(\half\,\quarter\,\half)$, can be reconciled with theoretical calculations performed for the electron-doped phase with the \textit{A}$_x$Fe$_2$Se$_2$ stoichiometry (not matching the average chemical composition of the sample) under the assumption of a $d$-wave symmetry of the SC order parameter \cite{Maier11, DasBalatsky11}. Alternatively, the metallic phase could possibly be associated with (i) a vacancy-disordered structure \cite{WangSong11, HanShen11}, which would effectively result in a rigid-band shift and broadening of the \textit{A}$_x$Fe$_2$Se$_2$ electronic bands \cite{CaoDai11}, (ii) an electron-doped \textit{A}$_x$Fe$_4$Se$_5$ phase \cite{DasBalatsky11a} with full or partial vacancy ordering, or (iii) possess a different $\sqrt{2}\times\!\sqrt{2}$ superstructure that corresponds to the \textit{A}$_x$Fe$_{1.5}$Se$_2$ composition \cite{WangSong11, KazakovAbakumov11, WangWang11}. However, to the best of our knowledge, first-principles calculations of the spin-excitation spectrum are not yet available for any of these alternative scenarios.

To be able to differentiate between the mentioned possibilities and thus try to verify the origin of the spin-excitation spectrum, we have performed a detailed study of the reciprocal-space structure and the dispersion of the previously reported resonant mode. We show that the resonant magnetic excitations in the SC state of Rb$_x$Fe$_{2-y}$Se$_2$ are quasi-two-dimensional (2D) and exhibit a complex in-plane pattern, dominated by four elliptical peaks that symmetrically surround the corner of the unfolded Brillouin zone (BZ) \cite{ParkInosov10}. This result is consistent with the dynamic spin susceptibility of an electron-doped \textit{A}$_x$Fe$_2$Se$_2$ compound, calculated in the SC state from a tight-binding model of the band structure by means of the random phase approximation (RPA) \cite{Maier11}.

The sample for this study is identical to the one used in Ref.\,\onlinecite{ParkFriemel11}. It comprises several coaligned single crystals with the chemical composition Rb$_{0.8}$Fe$_{1.6}$Se$_2$ and a total mass of $\sim\kern.5pt$1\,g. These crystals are bulk superconductors with a $T_\text{c}$ of $32$\,K \cite{ParkFriemel11}, which have been characterized by transport, magnetometry and specific heat measurements (batch BR16 in Ref.\,\onlinecite{Tsurkan11}). The experiments were conducted at the thermal-neutron spectrometer IN8 (ILL, Grenoble), which was operated both in the triple-axis-spectrometer (TAS) and in the \textit{FlatCone} multianalyzer configurations. The latter allowed us convenient mapping of the reciprocal space at a constant energy. In the \textit{FlatCone} configuration we utilized Si(111) monochromator and analyzer with the fixed final wave vector $k_\text{f}=3$\,\AA$^{-1}$. For the TAS measurements, we used pyrolytic graphite (002) monochromator and analyzer with double focusing. The TAS measurements were done with constant $k_\text{f}=2.662$\,\AA$^{-1}$ and 4.1\,\AA$^{-1}$, and a pyrolytic graphite filter was installed between the sample and the analyzer to suppress higher-harmonic contamination. In order to measure the dispersion and the intensity distribution of the resonant mode along the $\mathbf{c}$ axis, we mounted the sample in the $(2H\,H\,L)$ scattering plane. Subsequent investigations of in-plane excitations were done in the $(H\,K\,0)$ plane. Here and throughout the paper our notation is given in unfolded reciprocal lattice units (r.l.u.), which refer to the iron sublattice with the lattice parameters $a=b=2.76$\,{\AA} and $c=7.25$\,{\AA} \cite{ParkFriemel11}. The existence of an Fe-vacancy-ordered AFM phase in our sample has been verified by measuring the magnetic superstructure reflections \cite{ParkFriemel11}.

\begin{figure}[t]
\includegraphics[width=1.015\columnwidth]{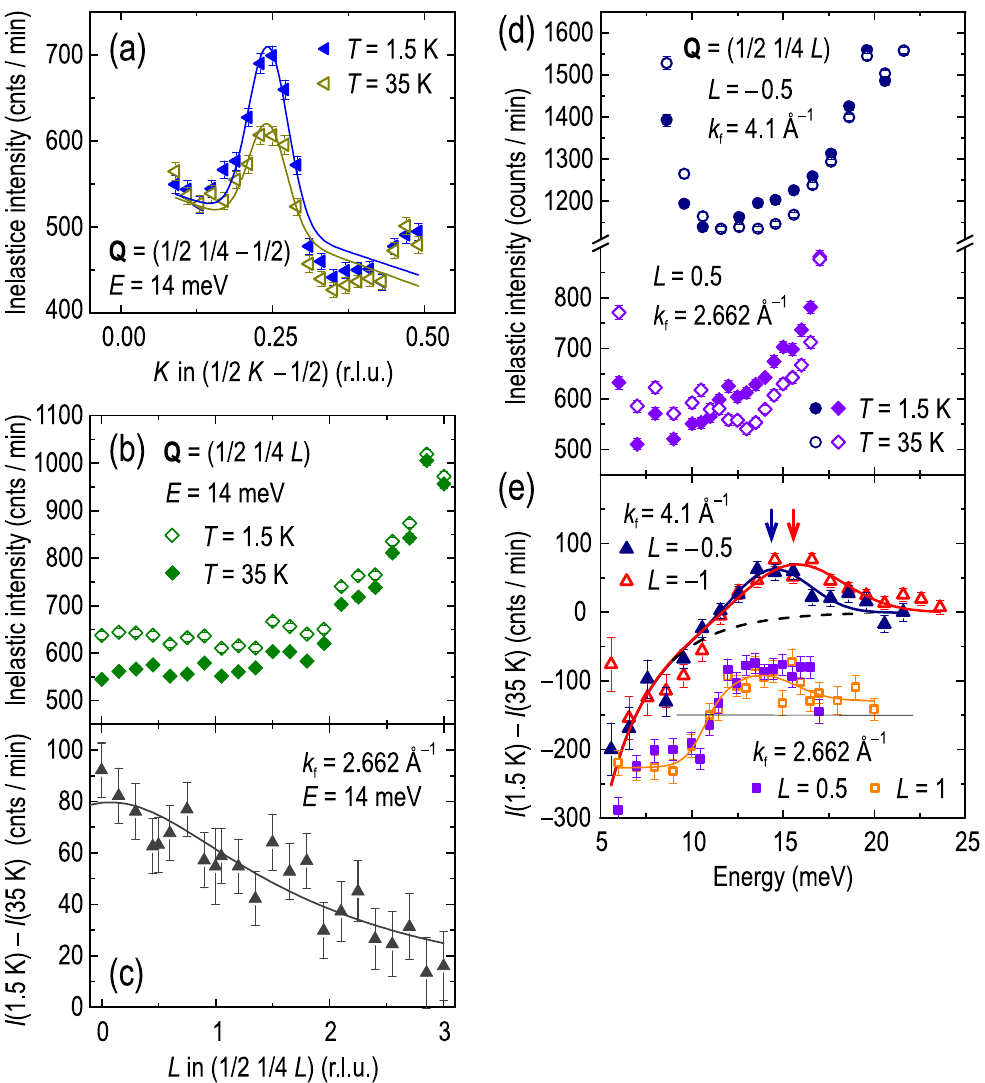}
\caption{(a)~Longitudinal momentum profiles through $\vect{Q}=(\half\,\quarter\,-\!\half)$ in the normal state at $35$\,K and in the SC state at $1.5$\,K, both at $E=14$\,meV. (b)~$L$-dependence of the intensity at $E=14$\,meV along $\vect{Q}=(\half\,\quarter\,L)$ at the same temperatures. (c)~The respective difference of the two signals fitted to the Fe$^{2+}$ magnetic form factor (solid line). (d)~Energy scans at the resonance position, measured at $L=-0.5$ with $k_\text{f}=4.1$\,\AA$^{-1}$ and at $L=0.5$ with $k_\text{f}=2.662$\,\AA$^{-1}$ in the normal and in the SC states. (e)~Difference of the SC- and normal-state intensities at integer and half-integer $L$. The solid line through the data points for $k_\text{f}=4.1$\,\AA$^{-1}$ is a fit with a Gaussian function superposed on the difference of the Bose factors for both temperatures (dashed line). The data points taken with $k_\text{f}=2.662$\,\AA$^{-1}$ are shifted down by 150 counts for clarity.\vspace{-1.5em}}
\label{fig:ldep}
\end{figure}

In Fig.\,\ref{fig:ldep}\,(a) we present longitudinal [as seen from $(\half\,\half\,0)$] momentum scans at the resonance energy, $\hslash\omega_{\rm res}=14$\,meV, along the $(\half~K -\!\half)$ direction in the normal state at 35\,K and in the SC state at $1.5$\,K. Already in the normal state, we observe a substantial magnetic response, which becomes considerably enhanced below $T_{\rm c}$. The center of both peaks lies at $K_0=(0.244\pm0.002)$, close to the commensurate position at $K=\quarter$. In Fig.\,1\,(b), the INS intensity at $\vect{Q}=(\half\,\quarter)$ and $E=14$\,meV is plotted as a function of the out-of-plane momentum component along $(\half\,\quarter\,L)$ for both temperatures. The intensity difference between 1.5 and 35\,K [Fig.\,\ref{fig:ldep}\,(c)], representing the resonant enhancement, is maximized at $L=0$ and then monotonically decreases for larger $L$ following the $\text{Fe}^{2+}$ magnetic form factor. This closely resembles the 2D nature of the signal in overdoped BaFe$_{2-x}$Co$_x$As$_2$ \cite{LumsdenChristianson09}, but is in contrast to the behavior of underdoped BaFe$_{2-x}$Ni$_x$As$_2$, where it is modulated as a function of $L$ and exhibits a maximum at $L=\half$ \cite{ParkInosov10}.

\begin{figure}[b]
\includegraphics[width=\columnwidth]{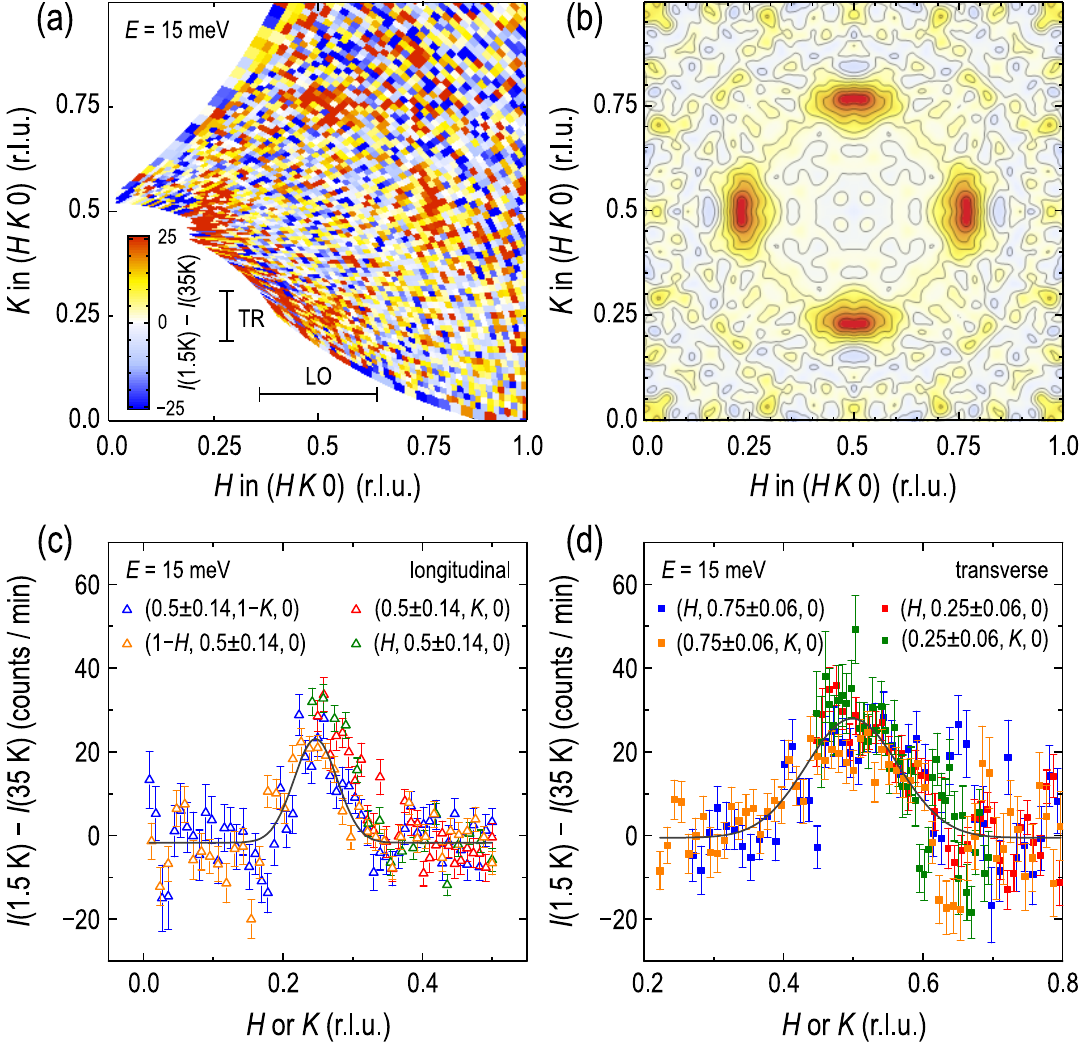}
\caption{(a)~Color map of the reciprocal space, showing intensity difference between the SC and normal states at $E=15$\,meV, measured by the \textit{FlatCone} detector. (b)~The same map as in (a), rebinned on a $81\times81$ grid, symmetrized with respect to the mirror planes and smoothed using a Gaussian filter with 1 pixel standard deviation. (c)~Longitudinal cuts (along the short axis of the ellipse) through the data in (a) at all four resonance positions. The intensity was integrated within a window of 0.28 r.l.u. in the direction perpendicular to the cut. (d)~The same in the transverse direction (long axis of the ellipse). The intensity was integrated within a window of 0.12 r.l.u. in the direction perpendicular to the cut. The widths of the integration windows are given by the horizontal and vertical bars in panel (a), respectively.\vspace{-1.3em}}
\label{fig:flatc}
\end{figure}

The weak $L$-dependence of the resonant signal can also be inferred from the energy scans measured above and below $T_{\rm c}$, such as those presented in Fig.\,\ref{fig:ldep}\,(d) for $k_{\rm f}=2.662$\,\AA$^{-1}$ and $4.1$\,\AA$^{-1}$. The intensity difference between the two temperatures is shown in panel (e) for both integer and half-integer $L$. Neither the energy nor the amplitude of the resonance peak shows any notable $L$-dependence beyond the uncertainty of the fit. This fact is consistent with ARPES measurements of the weak $k_\text{z}$-dispersion of the electron band at the $M$ point and of the SC gap \cite{Gap2}.

Based on the quasi-2D character of the magnetic intensity, we have mapped out the resonant enhancement of spin excitations at $E=15$\,meV in the $(HK0)$ scattering plane by means of the \textit{FlatCone} multianalyzer. Figure \ref{fig:flatc} (a) shows the difference of intensity maps measured around the BZ corner in the SC and normal states. We observe resonant intensity at all four symmetric positions equivalent to $(\half~\quarter\,0)$. In order to reduce the statistical noise in the data, we have rebinned this data set on an 81$\times$81 grid and symmetrized it with respect to four mirror planes of the reciprocal space, with subsequent Gaussian smoothing. The resulting intensity map is shown in Fig.\,\ref{fig:flatc} (b) as a contour plot.

One sees that the in-plane shape of the resonant intensity takes an elliptical form, elongated transversely with respect to the vector connecting it to $(\half\,\half\,0)$. We emphasize this by presenting cuts through all ellipses in the map of Fig.\,\ref{fig:flatc}\,(a) in the longitudinal [Fig.\,\ref{fig:flatc}\,(c)] and transverse [Fig.\,\ref{fig:flatc}\,(d)] directions. The intensity is integrated over the whole extension of the ellipse perpendicular to the cut, as indicated by the black bars in (a), in order to capture the whole resonant intensity. We observe an agreement between equivalent cuts with the same orientation. The ratio of the peak widths in the transverse and longitudinal directions results in an aspect ratio of $2.1$ for the resonance feature.

\begin{figure}[t]
\includegraphics[width=\columnwidth]{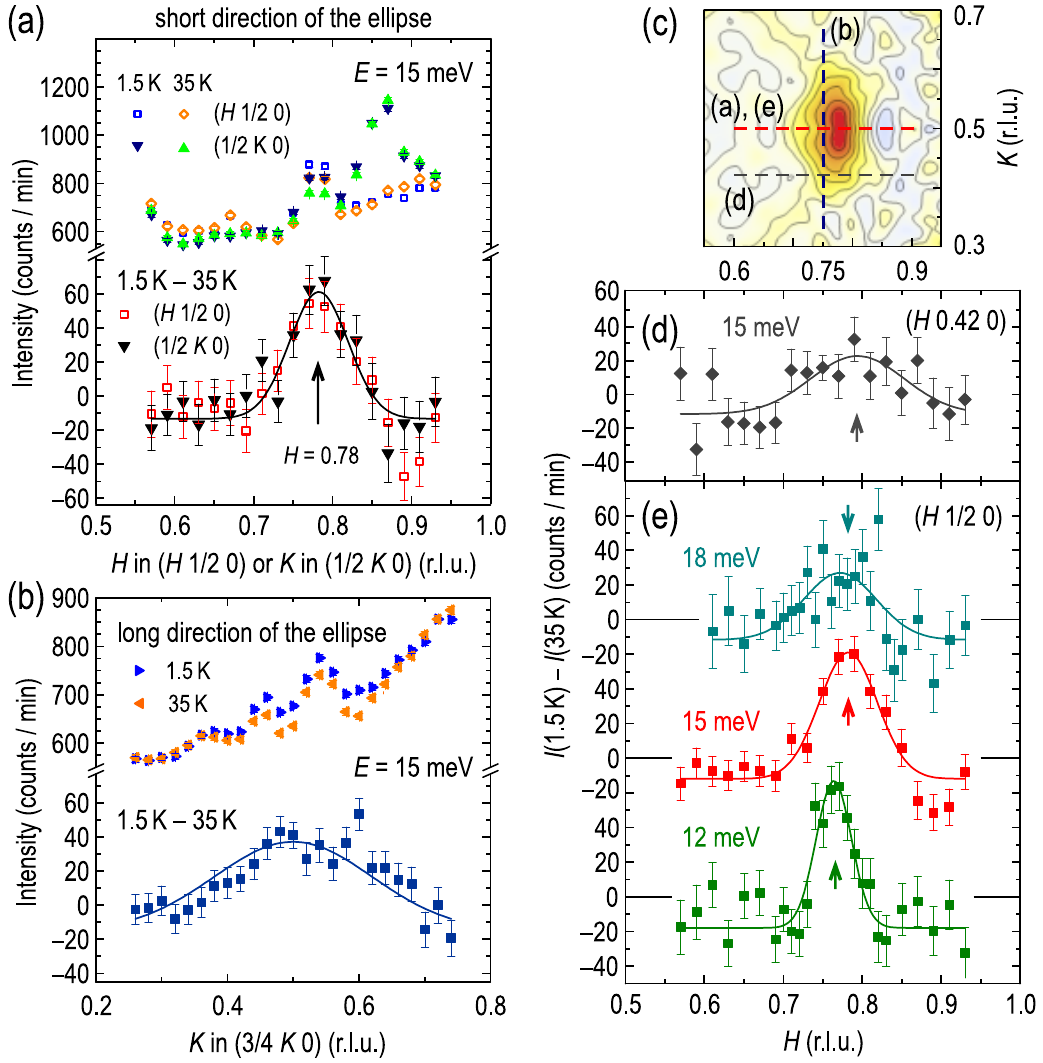}
\caption{TAS-mode data. (a)~Longitudinal momentum scans through the center of the ellipse at $\vect{Q}_1=(\quarterIII\,\half\,0)$ (triangles) and $\vect{Q}_2=(\half\,\quarterIII\,0)$ (squares and diamonds) as indicated in sketch (c) at $E=15$\,meV. The intensity in the SC and in the normal states (top) is shown together with their difference (bottom). (b)~The same for transverse momentum scans at $\vect{Q}_2=(\quarterIII\,\half\,0)$. (c)~A fragment of the \textit{FlatCone} map from Fig.\,\ref{fig:flatc} that illustrates the directions of the scans shown in this figure. Panels (d) and (e) show only the difference in intensity between SC and normal states. (d)~Momentum scan at $E=15$\,meV parallel to the longitudinal direction at $K=0.42$, offset from the center of the ellipse. (e)~Momentum scans at different energies along the short axis of the ellipse. The plot at $E=15$\,meV is an average of the two profiles in panel (a) at both resonance positions.\vspace{-1.4em}}
\label{fig:TAS}
\end{figure}

Next, we present momentum scans measured by TAS in longitudinal [Fig.\,\ref{fig:TAS}\,(a)] and transverse [Fig.\,\ref{fig:TAS}\,(b)] directions through the ellipse in the SC and normal states. Again, strongly anisotropic widths of the transverse and longitudinal profiles are observed in the intensity difference. The peak in the longitudinal direction for both resonances near $(\quarterIII\,\half\,0)$ and $(\half\,\quarterIII\,0)$ in Fig.\,\ref{fig:TAS}\,(a) is found at an incommensurate position of $H=0.78$ or $K=0.78$, respectively, as marked by the arrow. This is also consistent with the \textit{FlatCone} data in Fig.\,\ref{fig:TAS}\,(c), where the peak intensity is offset to the right from $H=\quarterIII$.

An elliptical in-plane shape of the resonance has also been observed in BaFe$_{2-x}$Co$_x$As$_2$ \cite{InosovPark10, ParkInosov10} and in Ba$_{1-x}$K$_x$Fe$_2$As$_2$ \cite{ZhangWang11} at the BZ boundary, so that both axes of the ellipse are aligned along the natural mirror planes of the reciprocal space. In Rb$_x$Fe$_{2-y}$Se$_2$, however, the ellipse could be asymmetric, because $H=\quarterIII$ is not a natural high-symmetry plane. Indeed, the shape in Fig.\,\ref{fig:TAS}\,(c) suggests a slight bending of the ellipse towards $(\half\,\half\,0)$. In the colormap in Fig.\,\ref{fig:TAS}\,(c), we also observe weak streaks of intensity reaching towards $(\half\,\quarter\,0)$ and $(\quarterIII\,\half\,0)$, barely above the statistical noise level, which could form parts of a ring connecting all four resonance positions. Nevertheless, the peak profile measured parallel to the longitudinal direction and offset by $0.08\,\text{r.l.u.}$ from the center of the ellipse [Fig.\,\ref{fig:TAS}\,(d)] does not show any notable shift of the peak center beyond statistical uncertainty. This indicates a nearly symmetric (non-curved) shape of the resonance peak in the vicinity of its maximum.

Finally, we turn to the in-plane dispersion of the resonance, which could be studied due to the broad distribution of the resonant intensity in energy, as can be seen in Fig.\,1\,(e). Figure~\ref{fig:TAS}\,(e) presents longitudinal momentum scans of the resonant intensity at 12, 15, and 18\,meV. Here, the peak center shifts from $H=(0.764\pm0.002)$\,r.l.u. at 12\,meV to $H=(0.782\pm0.003)$\,r.l.u. at 15\,meV, although we do not resolve a further shift upon changing the energy to 18\,meV. Moreover, comparison of the peak position at $L=-0.5$ [Fig.\,\ref{fig:ldep}\,(a)], centered at $H=0.244\pm0.002$, and at $L=0$ [Fig.\,\ref{fig:TAS}\,(a)], where it is shifted to a position equivalent to $H=0.218\pm0.003$, also suggests a small ($\sim$\,10\%) variation in the peak position along the $\mathbf{c}$ axis.

\begin{figure}[t]
\includegraphics[width=\columnwidth]{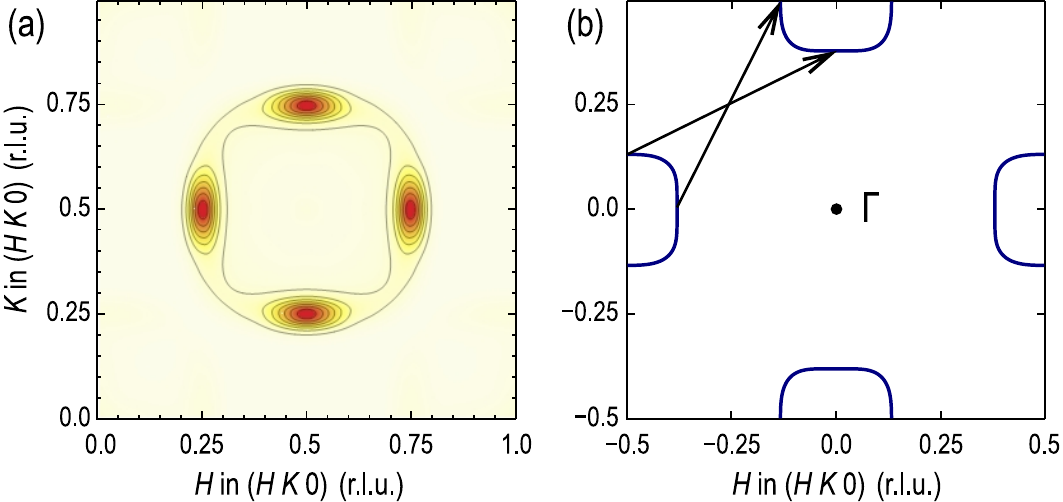}
\caption{(a) The difference of the calculated imaginary parts of the dynamic spin susceptibility for the SC and normal states, taken at the resonance energy, $\chi''_{\rm SC}(\vect{Q},\omega_{\rm res})-\chi''_{\rm n}(\vect{Q},\omega_{\rm res})$. The calculation was done within RPA from the tight-binding band model of \textit{A}$_x$Fe$_2$Se$_2$ \cite{Maier11}, which was rigidly shifted to match the experimental peak positions. An isotropic Gaussian broadening with a standard deviation of 0.02~r.l.u. has been applied to mimic the experimental resolution. (b) The resulting FS in the $(H\,K\,0)$ plane corresponds to the doping level of 0.18 electrons/Fe. The black arrows are the in-plane nesting vectors responsible for the resonance peaks observed in our present study.\vspace{-1.3em}}
\label{fig:calc}
\end{figure}

To verify the origin of the observed spectrum of spin excitations in Rb$_x$Fe$_{2-y}$Se$_2$, we will now compare our experimental observations with the results of band structure calculations. For this purpose, we employ the tight-binding model that was introduced in Ref.\,\onlinecite{Maier11} to describe the electronic structure of an electron-doped \textit{A}$_x$Fe$_2$Se$_2$. The chemical potential has been adjusted by a rigid-band shift of the bands to match the positions of the magnetic resonant peaks in the calculated susceptibility with the experimental data. This resulted in a doping level of $\sim$\,0.18 electrons/Fe. To enable direct comparison between the theory and experiment, we have calculated the imaginary part of the dynamical spin susceptibility at the resonance energy, $\chi''(\vect{Q},\omega_{\rm res})$, both for the SC and the normal states, as described in Ref.~\onlinecite{Maier11}. For the calculation in the SC state, we have assumed a $d_{x^2-y^2}$ gap $\Delta(\mathbf{k})=\Delta_0(\cos k_x-\cos k_y)$ \cite{note}. The color map in Fig.\,\ref{fig:calc}\,(a) shows the respective difference of the two quantities, $\chi''_{\rm SC}(\vect{Q},\omega_{\rm res})-\chi''_{\rm n}(\vect{Q},\omega_{\rm res})$, within the $(H\,K\,0)$ plane, isotropically broadened by a Gaussian resolution function with a standard deviation of 0.02~r.l.u. Comparison with the experimentally measured resonant intensity map in Fig.\,\ref{fig:flatc}\,(b) reveals good agreement between the two $\vect{Q}$-space patterns, as both the orientation and the aspect ratio of the elliptical peaks is well captured by the calculation. The origin of these peaks can be traced back to the nesting of electronlike Fermi pockets, as indicated in Fig.\,\ref{fig:calc}\,(b) by black arrows.

To conclude, the fact that the complicated pattern of resonant intensity in $\vect{Q}$-space can be successfully reproduced by our calculation strongly supports the itinerant origin of the observed magnetic response. The signal shows no signatures of the $\sqrt{5}\times\!\sqrt{5}$ reconstruction, indicating that it originates in the metallic phase of the sample without iron-vacancy ordering, as suggested recently \cite{WangSong11, SongWang11, LiDong11}. This distinguishes the newly observed signal from the previously reported spin-wave excitations in this class of compounds \cite{WangFang11} that stem from the magnetic superstructure Bragg positions in the insulating vacancy-ordered phase and are insensitive to the SC transition. The incommensurability of the resonance peak, as well as its variation with the out-of-plane momentum component and with energy, further indicates that it is not pinned to a particular position in $\vect{Q}$-space, but is arbitrarily determined by the level of electron doping, in line with the assumptions of Ref.\,\onlinecite{Maier11}.

Furthermore, we note that iron-pnictide compounds generally exhibit a tendency towards a 2D behavior of spin fluctuations with an increase of the doping level or $T_{\rm c}$ \cite{ParkInosov10, InosovPark11}. For example, optimally doped Ba$_{1-x}$K$_{x}$Fe$_2$As$_2$, which has the highest known critical temperature among all 122-compounds, shows almost no dispersion of the resonant energy, $\hslash\omega_{\rm res}$, along the $\mathbf{c}$ direction \cite{ZhangWang11}. Our data on Rb$_x$Fe$_{2-y}$Se$_2$ with a comparable transition temperature are fully consistent with this trend.
\enlargethispage{1em}

This work has been supported, in part, by the DFG within the Schwerpunktprogramm 1458, under Grants No. BO3537/1-1 and DE1762/1-1, and via TRR80 (Augsburg-Munich). T.~A.\,M. acknowledges support from the Center for Nanophase Materials Sciences, which is sponsored at Oak Ridge National Laboratory by the Scientific User Facilities Division, US Department of Energy. We thank S.\,Graser for providing the 5-band tight-binding model that was used in the calculation and acknowledge helpful discussions with A.~Boris, P.~Bourges, A.~Bosak, D.~Chernyshov, P.~J.~Hirschfeld, I.\,I.~Mazin, V.~Yu.~Pomjakushin, D.\,J.~Scalapino, Y.~Sidis, and A.~Yaresko.


\vspace{-1.0em}\vfill
\begin{thebibliography}{10}\vspace{-0.3em}\vfill
{\item[\hspace{1.7em}\large *]\hspace{-0.38em}Corresponding author: \href{mailto:d.inosov@fkf.mpg.de}{d.inosov@fkf.mpg.de}\label{CorrAuthor}\vspace{0.8em}}

\bibitem{Discovery} J.\,-G.\,Guo \textit{et~al.}, \href{http://link.aps.org/abstract/PRB/v82/e180520}{\prb {\bf 82}, 180520(R) (2010)};
                    A.\,F.\,Wang \textit{et~al.}, \href{http://link.aps.org/abstract/PRB/v83/e060512}{\textit{ibid}. {\bf 83}, 060512(R) (2011)};
                    A.\,Krzton-Maziopa \textit{et~al.}, \href{http://iopscience.iop.org/0953-8984/23/5/052203}{J.~Phys.:~Condens.\,Matter {\bf 23}, 052203 (2011)}.
\bibitem{Shermadini11} Z.\,Shermadini \textit{et~al.}, \href{http://link.aps.org/abstract/PRL/v106/e117602}{\prl {\bf 106}, 117602 (2011)}.
\bibitem{Yan11} R.\,H.\,Liu \textit{et al.}, \href{http://iopscience.iop.org/0295-5075/94/2/27008}{EPL {\bf 94}, 27008 (2011)}; Y.\,J.\,Yan \textit{et al.}, \href{http://arxiv.org/abs/1104.4941}{arXiv:1104.4941}.
\bibitem{order} V.~Y.~Pomjakushin \textit{et al.}, \href{http://link.aps.org/abstract/PRB/v83/e144410}{\prb {\bf 83}, 144410 (2011)};
                F.~Ye \textit{et al.}, \href{http://link.aps.org/doi/10.1103/PhysRevLett.107.137003}{\prl \textbf{107}, 137003 (2011)};
                W.~Bao \textit{et al.}, \href{http://cpl.iphy.ac.cn/qikan/epaper/zhaiyao.asp?bsid=12618}{Chin.~Phys.~Lett. {\bf 28}, 086104 (2011)}.
\bibitem{ZhangXiao11} A.~M.~Zhang \textit{et al.}, \href{http://arxiv.org/abs/1106.2706}{arXiv:1106.2706}.
\bibitem{Tsurkan11} V.~Tsurkan \textit{et al.}, \href{http://link.aps.org/abstract/PRB/v84/e144520}{\prb {\bf 84}, 144520 (2011)}.
\bibitem{stm} W.~Li \textit{et al.}, Nature Phys., \href{http://www.nature.com/nphys/journal/vaop/ncurrent/abs/nphys2155.html}{doi:10.1038/nphys2155} (in press); P.~Cai \textit{et al.}, \href{http://arxiv.org/abs/1108.2798}{arXiv:1108.2798}.
\bibitem{Charnukha11} A.\,Charnukha \textit{et al.}, \href{http://arxiv.org/abs/1108.5698}{arXiv:1108.5698};
                      C.\,C.\,Homes, Z.\,J. Xu, J.\,S. Wen, and G.\,D. Gu, \href{http://arxiv.org/abs/1110.5529}{arXiv:1110.5529}.
\bibitem{YanGao11} X.-W.~Yan, M.\,Gao, Z.-Y.~Lu, and T.~Xiang, \href{http://link.aps.org/abstract/PRB/v83/e233205}{\prb {\bf 83}, 233205 (2011)}.
\bibitem{CaoDai11} C.\,Cao and J.\,Dai \textit{et al.}, \href{http://prl.aps.org/abstract/PRL/v107/i5/e056401}{\prl \textbf{107}, 056401 (2011)}.
\bibitem{Gap1} X.-P.~Wang \textit{et al.}, \href{http://iopscience.iop.org/0295-5075/93/5/57001}{EPL {\bf 93}, 57001 (2011)};
              L.\,Zhao \textit{et al.}, \href{http://link.aps.org/abstract/PRB/v83/e140508}{\prb {\bf 83}, 140508(R) (2011)};
              D.\,Mou \textit{et al.}, \href{http://prl.aps.org/abstract/PRL/v106/i11/e107001}{\prl {\bf 106}, 107001 (2011)};
              T.~Qian \textit{et al.}, \href{http://prl.aps.org/abstract/PRL/v106/i18/e187001}{\textit{ibid.} {\bf 106}, 187001 (2011)}.
\bibitem{Gap2} Y.\,Zhang \textit{et al.}, \href{http://www.nature.com/nmat/journal/v10/n4/full/nmat2981.html}{Nature Mater. {\bf 10}, 273 (2011)}.
\bibitem{WangSong11} Z.\,Wang \textit{et al.}, \href{http://prb.aps.org/abstract/PRB/v83/i14/e140505}{\prb \textbf{83}, 140505 (2011)}.
\bibitem{SongWang11} Y.\,J.\,Song \textit{et al.}, \href{http://iopscience.iop.org/0295-5075/95/3/37007/}{EPL \textbf{95}, 37007 (2011)}.
\bibitem{YuanDong11} R.\,H.\,Yuan \textit{et al.}, \href{http://arxiv.org/abs/1102.1381}{arXiv:1102.1381}.
\bibitem{KazakovAbakumov11} S.\,M.\,Kazakov \textit{et al.}, \href{http://pubs.acs.org/doi/abs/10.1021/cm201203h}{Chem. Mater. {\bf 23}, 4311 (2011)}.
\bibitem{xray} A.\,Ricci \textit{et al.}, \href{http://prb.aps.org/abstract/PRB/v84/i6/e060511}{\prb \textbf{84}, 060511 (2011)}.
\bibitem{ChenXu11} F.\,Chen \textit{et al.}, \href{http://arxiv.org/abs/1106.3026}{arXiv:1106.3026}.
\bibitem{ShenZeng11} B.\,Shen \textit{et al.}, \href{http://epljournal.edpsciences.org/articles/epl/abs/2011/21/epl13941/epl13941.html}{EPL \textbf{96}, 37010 (2011)}.
\bibitem{ShermadiniLuetkens11} Z.\,Shermadini \textit{et al.}, \href{http://arxiv.org/abs/1111.5142}{arXiv:1111.5142}.
\bibitem{KsenofontovWortmann11} V.\,Ksenofontov \textit{et al.}, \href{http://link.aps.org/abstract/PRB/v84/e180508}{\prb {\bf 84}, 180508(R) (2011)}.
\bibitem{spm} I.~I.~Mazin, D.~J.~Singh, M.~D.~Johannes, and M.~H.~Du, \href{http://prl.aps.org/abstract/PRL/v101/i11/e057003}{\prl {\bf 101}, 057003 (2008)};
    K.\,Kuroki \textit{et al.}, \href{http://prl.aps.org/abstract/PRL/v101/i8/e087004}{\textit{ibid.} {\bf 101}, 087004 (2008)};
    A.\,D.\,Christianson \textit{et al.}, \href{http://www.nature.com/nature/journal/v456/n7224/full/nature07625.html}{Nature (London) {\bf 456}, 930 (2008)};
    C.-T. Chen, C.\,C. Tsuei, M.\,B. Ketchen, Z.-A. Ren, and Z.\,X. Zhao, \href{http://www.nature.com/nphys/journal/v6/n4/full/nphys1531.html}{Nature Phys. {\bf 6}, 260 (2010)}.
\bibitem{LumsdenChristianson09} M.\,D.\,Lumsden \textit{et al.}, \href{http://prl.aps.org/abstract/PRL/v102/i10/e107005}{\prl \textbf{102}, 107005 (2009)}.
\bibitem{InosovPark10} D.\,S.\,Inosov \textit{et al.}, \href{http://www.nature.com/nphys/journal/v6/n3/full/nphys1483.html}{Nature Phys. {\bf 6}, 178 (2010)}.
\bibitem{LDA} X.-W.~Yan, M.\,Gao, Z.-Y.~Lu, and T.~Xiang, \href{http://link.aps.org/abstract/PRB/v84/e054502}{\prb {\bf 84}, 054502 (2011)};
              C.\,Cao and J.\,Dai, \href{http://cpl.iphy.ac.cn/qikan/epaper/zhaiyao.asp?bsid=12354}{Chin.~Phys.~Lett. {\bf 28}, 057402 (2011)};
              I.\,R.\,Shein and A.\,L.\,Ivanovskii, \href{http://www.sciencedirect.com/science/article/pii/S037596011001635X}{Phys.~Lett.~A {\bf 375}, 1028 (2011)};
              I.\,A.\,Nekrasov and M.\,V.~Sadovskii, \href{http://www.springerlink.com/content/p968541148465474/}{JETP~Lett. {\bf 93}, 166 (2011)}.
\bibitem{Mazin11} I.\,I.\,Mazin \textit{et al.}, \href{http://link.aps.org/doi/10.1103/PhysRevB.84.024529}{\prb \textbf{84}, 024529 (2011)}.
\bibitem{Maier11} T.\,A.\,Maier, S.\,Graser, P.~J.\,Hirschfeld, and D.\,J.\,Scalapino, \href{http://link.aps.org/abstract/PRB/v83/e100515}{\prb {\bf 83}, 100515(R) (2011)}.
\bibitem{DasBalatsky11} T.~Das and A.\,V.~Balatsky, \href{http://prb.aps.org/abstract/PRB/v84/i1/e014521}{\prb \textbf{84}, 014521 (2011)}.
\bibitem{ParkFriemel11} J.\,T.~Park \textit{et al.}, \href{http://link.aps.org/abstract/PRL/v107/e177005}{\prl \textbf{107}, 177005 (2011)}.
\bibitem{HanShen11} F.~Han, B.\,Shen, Z.-Y.~Wang, and H.-H.\,Wen, \href{http://arxiv.org/abs/1103.1347}{arXiv:1103.1347}.
\bibitem{DasBalatsky11a} T.~Das and A.\,V.~Balatsky, \href{http://link.aps.org/abstract/PRB/v84/e115117}{\prb \textbf{84}, 115117 (2011)}.
\bibitem{WangWang11} M.\,Wang \textit{et al.}, \href{http://link.aps.org/doi/10.1103/PhysRevB.84.094504}{\prb \textbf{84}, 094504 (2011)}.
\bibitem{ParkInosov10} J.\,T.~Park \textit{et al.}, \href{http://prb.aps.org/abstract/PRB/v82/i13/e134503}{\prb \textbf{82}, 134503 (2010)}.
\bibitem{ZhangWang11} C.\,Zhang \textit{et al.}, \href{http://www.nature.com/srep/2011/111019/srep00115/full/srep00115.html}{Scientific Reports \textbf{1}, 115 (2011)}.
\bibitem{note} The purpose of this assumption is to obtain different signs of the order parameter on the two FS sheets, which could also be achieved with alternative pairing symmetries \cite{Mazin11} without significantly affecting our final result.
\bibitem{LiDong11} W.~Li, S.\,Dong, C.\,Fang, and J.\,Hu, \href{http://arxiv.org/abs/1110.0372}{arXiv:1110.0372}.
\bibitem{WangFang11} M.\,Wang \textit{et al.}, \href{http://www.nature.com/ncomms/journal/v2/n12/abs/ncomms1573.html}{Nature Commun. {\bf 2}, 580 (2011)}.
\bibitem{InosovPark11} D.\,S.\,Inosov \textit{et al.}, \href{http://link.aps.org/doi/10.1103/PhysRevB.83.214520}{\prb \textbf{83}, 214520 (2011)}.
\end{thebibliography}
\end{document}